# Explainable Artificial Intelligence for Manufacturing Cost Estimation and Machining Feature Visualization

Soyoung Yoo
Department of Mechanical Systems Engineering,
Sookmyung Women's University, 04310, Seoul, Korea
ysy@sm.ac.kr

Namwoo Kang*
The Cho Chun Shik Graduate School of Green Transportation,
Korea Advanced Institute of Science and Technology, 34141, Daejeon, Korea
nwkang@kaist.ac.kr

*Corresponding author

## Abstract

Studies on manufacturing cost prediction based on deep learning have begun in recent years, but the cost prediction rationale cannot be explained because the models are still used as a black box. This study aims to propose a manufacturing cost prediction process for 3D computer-aided design (CAD) models using explainable artificial intelligence. The proposed process can visualize the machining features of the 3D CAD model that are influencing the increase in manufacturing costs. The proposed process consists of (1) data collection and pre-processing, (2) 3D deep learning architecture exploration, and (3) visualization to explain the prediction results. The proposed deep learning model shows high predictability of manufacturing cost for the computer numerical control (CNC) machined parts. In particular, using 3D gradient-weighted class activation mapping proves that the proposed model not only can detect the CNC machining features but also can differentiate the machining difficulty for the same feature. Using the proposed process, we can provide a design guidance to engineering designers in reducing manufacturing costs during the conceptual design phase. We can also provide real-time quotations and redesign proposals to online manufacturing platform customers.

**Keywords**: Cost estimation, Machining feature, Deep learning, Explainable artificial intelligence (XAI), 3D CAD, 3D Grad-CAM

# 1. Introduction

Online manufacturing platforms are increasing, and they provide a service of manufacturing and shipping after customers upload their computer-aided design (CAD) models to a website (Wu et al., 2015). Estimating the manufacturing cost to provide a quotation to the customer takes time and money because the CAD model should be delivered to the manufacturer and then the manufacturing experts review the model. However, the manufacturing cost of a CAD model for on-demand service needs to be predicted in real time and feedback needs to be provided to the customer.

The manufacturing industry is always searching for ways to reduce manufacturing costs. Manufacturing costs are mainly estimated based on CAD drawings. Although the engineering design accounts for 5% of total cost, it determines 70% of the final manufacturing cost (Boothroyd, 1994). If manufacturing costs can be accurately predicted in real time in the early stages of development, then designs can be modified to achieve high engineering performance while meeting target costs.

Many manufacturing cost estimating studies have been conducted. Niazi et al. (2006) categorized the overall research into qualitative and quantitative techniques. The qualitative techniques are further subdivided into intuitive and analogical techniques and the quantitative ones into parametric and analytical techniques. Analogical methodology is a data-driven approach that predicts manufacturing costs based on historical product data, and regression analysis and artificial neural networks (ANNs) have been mainly used among these methodologies (Duran et al., 2012; Chan et al., 2018).

Deep learning research on estimating manufacturing costs has increased in recent years. For example, Ning et al. (2020) showed that 3D CAD models can be trained using a 3D convolutional neural network (CNN) to predict manufacturing costs. However, the deep learning models presented in previous studies have limitations in that they cannot explain the reason for determining manufacturing cost. As artificial intelligence (AI) technology develops, the importance of research on explainable AI (XAI) that can explain the rationale of AI judgment is increasing. Which feature of 3D CAD model mainly affects the increase in manufacturing costs needs to be explained for realiable prediction of manufacturing cost.

Therefore, the motivation for our study is to develop an AI model that informs designers which parts need to be modified in 3D CAD to reduce manufacturing costs. Although designers do not have domain knowledge of manufacturing costs, this AI model can provide guidance to designers in the early stages of product development to redesign their products to meet their target costs.

This research aims to present a deep learning-based manufacturing cost estimation and machining feature visualization process for explaining the rationale of estimated manufacturing cost. First, a 3D CNN model using voxelized 3D CAD, materials, and volume data as inputs is built. Second, 3D CAD features that affect manufacturing costs are visualized through 3D gradient-weight class activation mapping (Grad-CAM) to explain the reason of judgment. This study demonstrates the performance of the proposed process by using 3D CAD models and price data of the computer numerical control (CNC) machined parts being sold online. The results show that the proposed model not only can detect the CNC machining features but also can differentiate the machining difficulty.

The rest of the paper is organized as follows. Section 2 introduces related works. Section 3 introduces the proposed framework. Section 4 describes the cost prediction results and shows machining feature visualization results. Section 5 summarizes conclusions and future research directions.

# 2. Literature Review

Section 2.1 introduces previous cost estimation studies, and Section 2.2 introduces 3D deep learning architectures using 3D CAD benchmarked in this work.

## 2.1. Cost Estimation

On the basis of the classification of cost estimation methodology by Niazi et al. (2006), this study focuses on the analogical methodology, which is a data-driven approach for manufacturing cost estimation. In previous research, the analogical technique can be largely divided into the regression analysis model and the ANN model.

Rickenbacher et al. (2013) developed a cost model that can estimate the actual cost of a single part to integrate the additive manufacturing into the production process. The cost model uses linear regression to estimate manufacturing times for 24 different manufacturing operations. Chan et al. (2018) developed a cost estimation framework to estimate the manufacturing costs associated with new products based on previous products of similar shape. The proposed framework is implemented for additive manufacturing and applies machine learning algorithms for dynamic clustering, LASSO, and elastic net regressions.

Predictions through linear regression models can understand and interpret the effect of each variable. However, data with complex patterns are difficult to predict. One way to overcome these shortcomings is to use a nonlinear regression model. ANNs are mainly used as a nonlinear regression model.

Duran et al. (2012) developed and tested a model of manufacturing cost estimating of piping elements during the early design phase through the application of ANN. The developed model demonstrates that neural networks can improve the accuracy of cost estimation for shell and tube heat exchangers. Juszczyk (2017) applied the AI tools to construction, and it shows that the ANNs can be applied successfully on cost estimating on the level of construction works.

ANN model would allow obtaining an accurate prediction even when adequate information is unavailable in the early stages of the design process. However, ANNs cannot predict the cost based on 3D features because the input data are converted to 1D data and spatial information is not reflected. To overcome this limitation, Ning et al. (2020) proposed a 3D CNN model that can automatically extract features from 3D CAD data. The architecture of this model is presented in Fig. 1. However, this model cannot explain the rationale of manufacturing cost estimation and does not use information about the volume or material.

This study proposes a cost estimation process that can compensate for the shortcomings of the unexplainable deep learning-based cost estimation model. First, the areas that affect the cost resulting in voxelized 3D CAD can be visualized by applying the 3D Grad-CAM method to confirm the rationale of the manufacturing cost decision. Second, we use 3D CAD models and geometric volume and material data together to make more accurate cost estimates.

## 2.2. 3D Deep Learning Architectures

Various presentation methods of 3D data are available, such as primitive, projection, RGB-D, volume (voxel), multiview, point cloud, graph, and mesh (Ahmed et al., 2018). In computer science, the most widely used methods of expressing 3D data are voxels and point cloud; baseline deep learning models, such as VoxNet (Maturana and Scherer, 2015) and PointNet (Qi et al., 2017), are also available. VoxNet is used for as a base model for developing a 3D CNN-based architecture. To train a model, 3D data, such as point cloud CAD data, are voxelized for use as input data. For input data, feature extraction is performed by two 3D convolutional layers and one pooling layer. The output is then finally derived through two fully connected layers. However, voxelization unnecessarily increases the size of the data and causes a lot of computations. PointNet, which is an alternative to solving the problem, can learn point-wise features directly from point clouds. This approach demonstrates impressive results on 3D object recognition. PointNet can perform object part segmentation and object classification by using only the fully connected layer and max pooling without using 3D convolution.

Beyond computer science domain, deep learning studies using 3D CAD models are being actively conducted in engineering design and manufacturing area. Recently, Ning et al. (2020) introduced 3D CNN to predict manufacturing costs, as shown in Fig. 1(a). Dering and Tucker. (2017) presented 3D CNN-based model to classify the functions of 3D CAD, as shown in Fig. 1(b). Williams et al. (2019) proposed a 3D CNN model for additive manufacturing to predict the part mass, the support material mass, and the build time, as shown in Fig. 1(c). In our study, 3D CNN architectures of Ning et al. (2020), Dering and Tucker. (2017), and Williams et al. (2019) are used for baseline models to find the best architecture and compare the performances.

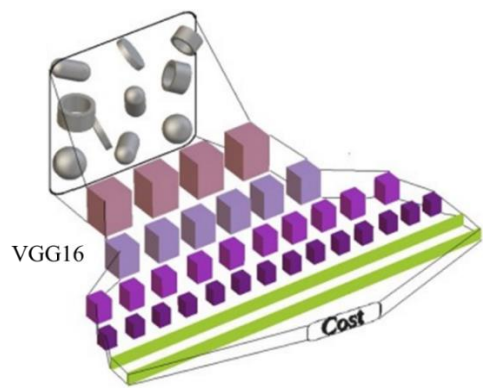

(a) 3D CNN arthitecture of Ning et al. (2020)

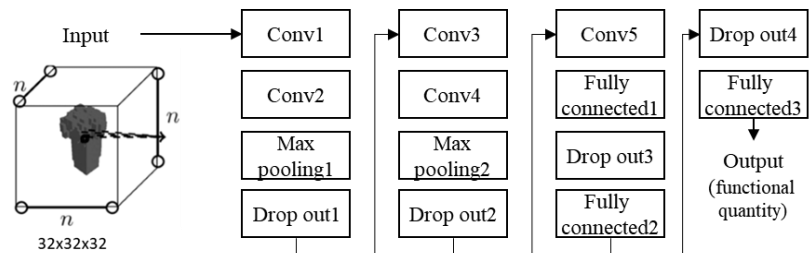

(b) 3D CNN arthitecture of Dering and Tucker. (2017)

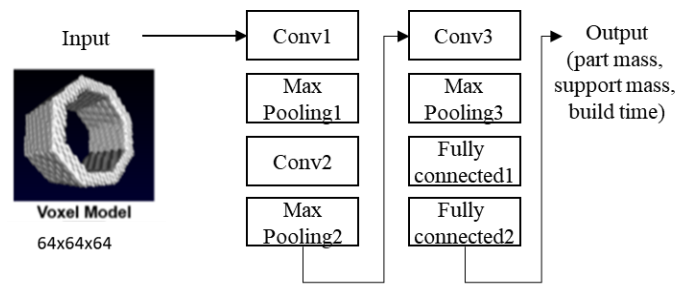

(c) 3D CNN arthitecture of Williams et al. (2019)

Fig. 1. Baseline architectures

CNC machined parts have machining features consisting of pockets, slots, and holes. Research has been actively conducted to automatically recognize these machining processing features, and FeatureNet (Zhang et al., 2018) is a method for detecting machining features using 3D CNN. The defined machining features from FeatureNet are used for test our proposed model in Section 5.1.

## 3. Research Framework

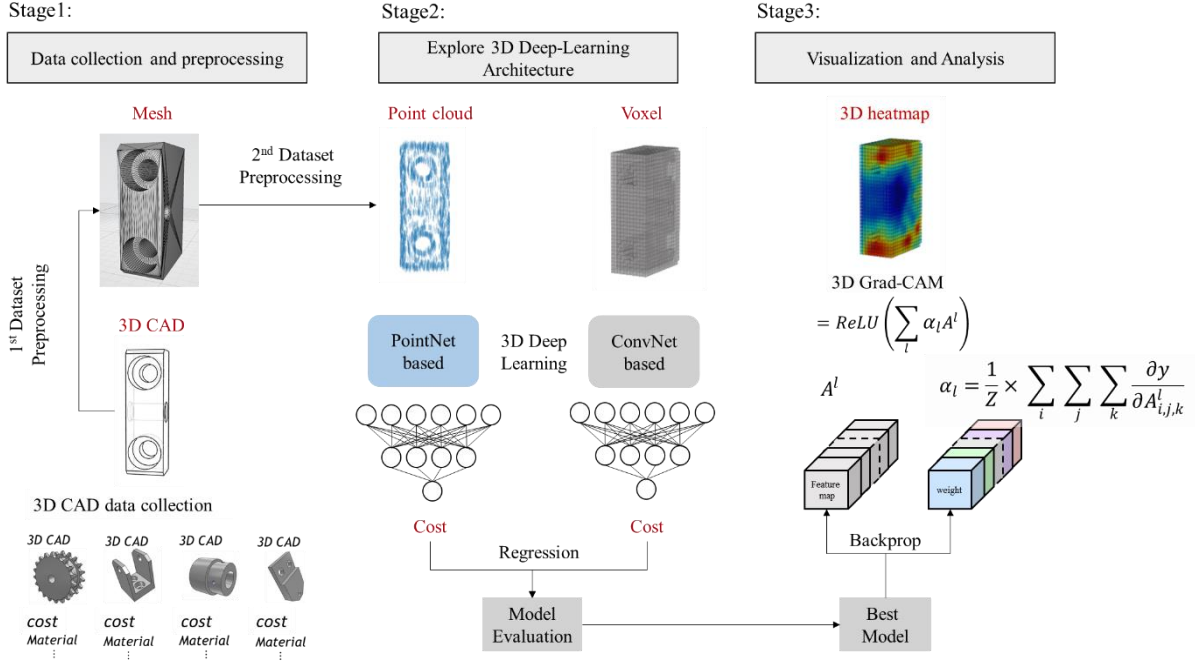

Fig. 2. Research Framework

The proposed framework of the manufacturing cost prediction is shown in Fig. 2, and each step is briefly discussed as follows.
- Stage 1: This stage collects 3D CAD, cost, and material information of 1,006 machined parts. Pre-processing of two stages should be performed to use the collected data as input to deep learning. In the first pre-processing, 3D CAD is converted to a mesh file, and the volume is calculated. Then, point cloud and voxel data are converted. In the second pre-processing, volume, material, and cost information are normalized. A detailed explanation of Stage 1 is given in Section 3.1.
- Stage 2: We evaluate the cost prediction performance of various architectures based on PointNet and 3D CNN to explore 3D deep learning architectures. We select the best model after evaluation. The proposed model is described in Section 3.2.2. All model evaluation results are described in Section 4.
- Stage3: 3D Grad-CAM is performed to explain the rationale of cost prediction result with visualization. We demonstrate the ability of the proposed model to detect CNC machining feature and distinguish machining difficulty. 3D Grad-CAM process is given in Section 3.3, and the visualization results are described in Section 5.

## 3.1 Data Collection and Pre-processing

This study collects 3D CAD data of CNC machined parts and associated price provided by MiSUMi (2020) online. In CNC machined parts, material costs generally account for more than 50% of the total manufacturing cost (Jung, 2002), and the manufacturing cost is lower when the difficulty of cutting is lower and the processing time is shorter. In our study, collected price data are used as manufacturing cost data under the assumption that manufacturing cost occupies a fixed proportion of price such that manufacturing cost and price has high correlation. A total of 1,006 data of parts are collected. A total of 34 categories of parts are considered, as shown in Fig. 3. 3D CAD, cost, and material data corresponding to each part are collected.

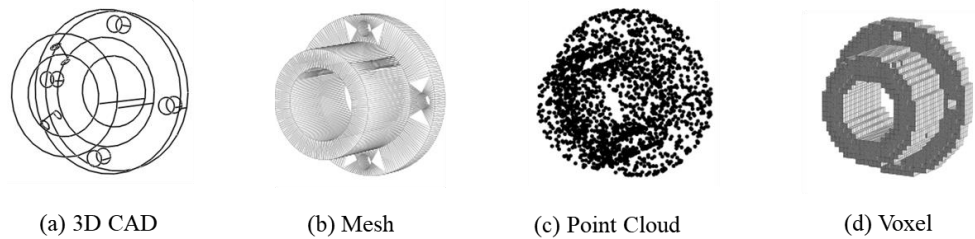

Fig. 3. Example of 3D CAD by category of collection machined parts

The collected data, which are 3D CAD and numerical data (volume, material, and cost), go through two pre-processing processes for use as a training dataset for deep learning models.

### 3.1.1. Pre-processing of 3D CAD

3D CAD pre-processing involves two stages: (1) mesh file conversion and (2) voxel and point cloud transformation. First, the collected 3D CAD is converted into a mesh file. Mesh files represent the shapes of 3D CAD models in numerous triangles and store triangular information as face and vertex information. In this study, FreeCAD (2020) with Python API are used to automate mesh file conversion. After converting to a mesh file, we use the Python library numpy-stl to obtain the volume of a 3D mesh shape.

Second, we create a 3D grid using the maximum and minimum values of the triangle's vertex coordinates (x, y, z) to convert mesh to voxel. Then, they are split into voxel grid sizes. In a divided 3D grid, a voxel is created in that grid when the triangle of the mesh file intersects in the grid. The voxelization algorithms used in this study use the methods presented by Adam (2020).

Third, the number of output points must be determined first to convert mesh to point cloud. Then, we use a weighted random sampling method to select triangles in the 3D mesh by $n$ points. Within each randomly selected triangle, one point in a random coordinate is created using the triangle's center of gravity method. Mesh is finally converted to point cloud after repeating this point-generation process for $n$ times. We use the point cloud transformation algorithm of Iglesia (2017). Fig. 4 shows examples of pre-processed 3D CAD data.

| (a) 3D CAD | (b) Mesh | (c) Point Cloud | (d) Voxel |

Fig. 4. Pre-processing 3D CAD data

### 3.1.2. Volume, Material, and Cost Pre-processing

The collected data are 3D cad data with cost, material, and volume. Data with large range deviations require a data scaling process before being used as inputs to deep learning models. Although various methods of data scaling are available, we use two techniques, namely, min–max normalization and log normalization, to pre-process the data as follows.

$$x_{new} = \frac{x - x_{min}}{x_{max} - x_{min}} \quad (4)$$

$$x_{new} = \ln x \quad (5)$$

The min–max normalization and the log normalization are expressed in Equations (4) and (5), respectively. Fig. 5 shows the results of data pre-processing with min–max normalization and log normalization. Fig. 5(a) is the result of scaling cost data, and Fig. 5(b) is the result of scaling volume data. The histogram in the first column is a visualization of the original data, and the histogram in the second column is a visualization of the data that are normalized to a min–max. The histogram in the last third column is the result of visualizing the data normalized to the log.

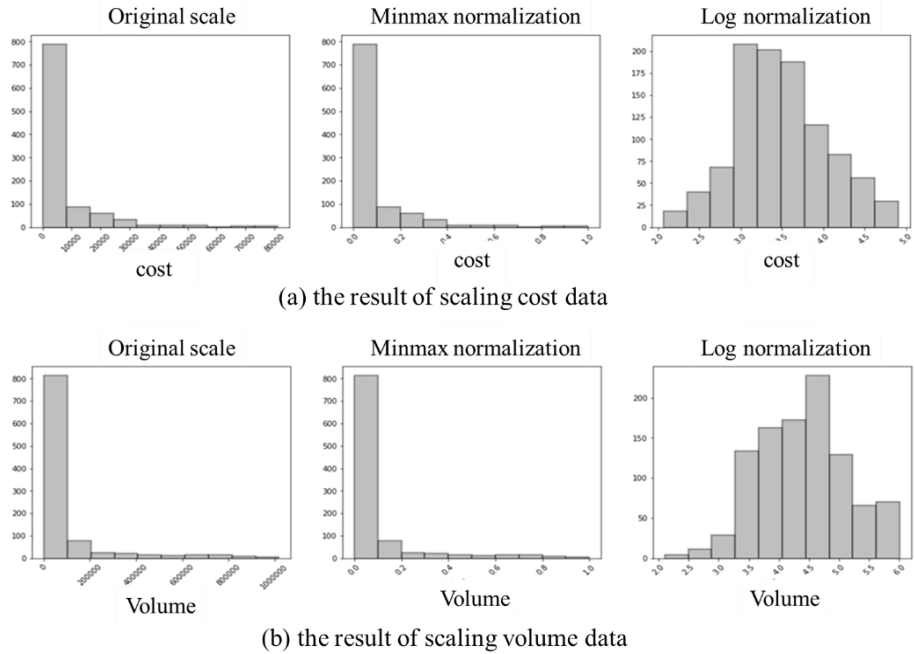

(a) the result of scaling cost data

(b) the result of scaling volume data

Fig. 5. Normalized cost and volume data histogram.

In our dataset, total parts are typically made of stainless steel and aluminum, and each representative material consists of a detailed material type. One-hot encoding method is used in this study given that detailed material data are categorical data. Table 1 shows the representative material of the whole part, each detailed material type, and one-hot vector of each material.

Table 1. Type of material for all parts

| Type | Detailed Type | One-hot vector |
|---|---|---|
| Steel | Structural Steel | [1,0,0,0,0,0,0,0,0,0,0,0,0] |
| Steel | S45C | [0,1,0,0,0,0,0,0,0,0,0,0,0] |
| Steel | S50C | [0,0,1,0,0,0,0,0,0,0,0,0,0] |
| Steel | SS400 | [0,0,0,1,0,0,0,0,0,0,0,0,0] |
| Steel | S35C | [0,0,0,0,1,0,0,0,0,0,0,0,0] |
| Aluminum | A6061 | [0,0,0,0,0,1,0,0,0,0,0,0,0] |
| Aluminum | Aluminum Alloys | [0,0,0,0,0,0,1,0,0,0,0,0,0] |
| Aluminum | A5052 | [0,0,0,0,0,0,0,1,0,0,0,0,0] |
| Aluminum | A2011 | [0,0,0,0,0,0,0,0,1,0,0,0,0] |
| Aluminum | 2000 Series Aluminum Alloys | [0,0,0,0,0,0,0,0,0,1,0,0,0] |
| Stainless | SUS304 | [0,0,0,0,0,0,0,0,0,0,1,0,0] |
| Stainless | Stainless | [0,0,0,0,0,0,0,0,0,0,0,1,0] |
| Stainless | SUS303 | [0,0,0,0,0,0,0,0,0,0,0,0,1] |

## 3.2. 3D Deep Learning Architecture

### 3.2.1 Baseline Models

We test deep learning models used in previous research to explore the best architecture and hyperparameters. This study benchmarks five models, namely, (1) VoxNet, (2) PointNet, (3) Dering and Tucker. (2017), (4) Williams et al. (2019), and (5) Ning et al. (2020), as introduced in Section 2.2. For the hyperparameter, we use 0.0001 for a learning rate and Adam for an optimizer. The batch size is set to 16, and the maximum epochs is used as 1000. However, early stopping techniques are used to prevent overfitting.

Model training is performed by varying 5 architectures, 2 pre-processing methods for cost and volume, 3 loss functions, and 3 combinations of input data as follows. As a result, 90 models (5×2×3×3=90) are tested as baseline models.

- **5 types of architecture**
  (1) VoxNet
  (2) PointNet
  (3) Dering and Tucker. (2017)
  (4) Williams et al. (2019)
  (5) Ning et al. (2020)
- **2 types of pre-processing method for cost and volume**
  (1) Min–max normalization;
  (2) Log normalization
- **3 types of loss function**
  (1) Mean Square Error (MSE)
  (2) Mean Absolute Error (MAE)
  (3) Mean Squared Logarithmic Error (MSLE)
- **3 types of input data combination**
  (1) 3D CAD
  (2) 3D CAD + Materials
  (3) 3D CAD + Materials + Volume

## 3.2.2 Proposed Model

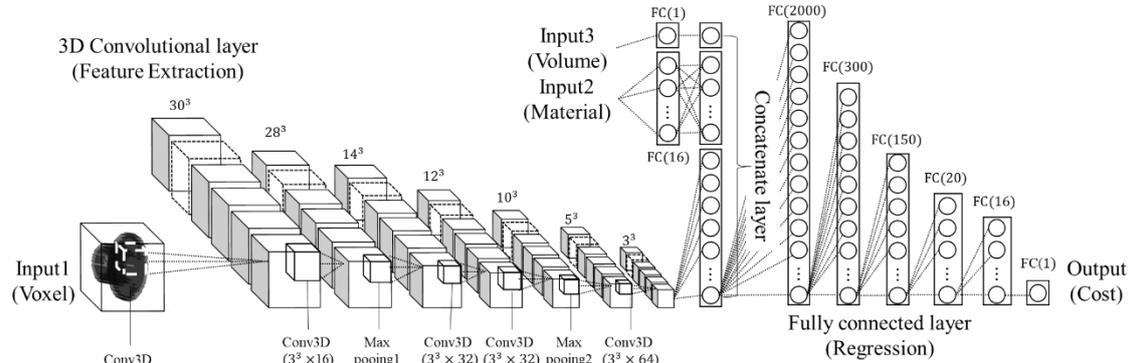

Fig. 6. Proposed model architecture

Table 2. Proposed model architecture

| Layer | Size (#number of filter) | Activation | Weight and bias initializer |
|---|---|---|---|
| Input1 | 32×32×32 | — | Xavier_normal, zeros |
| 3D Convolution | 3×3×3 (16) | LeakyReLU | Xavier_normal, zeros |
| 3D Convolution | 3×3×3 (16) | LeakyReLU | — |
| Max Pooling | 2×2×2 | — | — |
| Dropout | 0.3 | — | — |
| 3D Convolution | 3×3×3 (32) | LeakyReLU | Xavier_normal, zeros |
| 3D Convolution | 3×3×3 (32) | LeakyReLU | Xavier_normal, zeros |
| Max Pooling | 2×2×2 | — | — |
| Dropout | 0.3 | — | — |
| 3D Convolution | 3×3×3 (64) | LeakyReLU | Xavier_normal, zeros |
| Flatten (×1) | — | — | — |
| Input2 | 16 | — | — |
| Fully Connected (×2) | 16 | — | Xavier_normal, zeros |
| Input3 | 1 | — | — |
| Fully Connected (×3) | 1 | — | Xavier_normal, zeros |
| Concatenate layer (×1, ×2, ×3) | — | — | — |
| Fully Connected | 2000 | LeakyReLU | Xavier_normal, zeros |
| Fully Connected | 300 | LeakyReLU | Xavier_normal, zeros |
| Fully Connected | 150 | LeakyReLU | Xavier_normal, zeros |
| Fully Connected | 20 | LeakyReLU | Xavier_normal, zeros |
| Fully Connected | 16 | LeakyReLU | Xavier_normal, zeros |
| Fully Connected | 1 | — | — |

Fig. 6 and Table 2 show the architecture of the proposed model. The convolutional layers serve as feature extractions, and the fully connected layers are used for regression. We test various architectures adopted as baseline models and develop the proposed architecture with some modifications based on Dering and Tucker. (2017)'s architecture.

Dering and Tucker. (2017)'s model and the proposed model have differences. First, we adopt the feature extraction part, but we modify the regression part by making the layers deeper while reducing the number of neurons. Dering and Tucker. (2017) used only two hidden layers of 1024 neurons for regression part. However, the proposed model has 5 hidden layers, and the number of neurons decreased in the order of 2000, 300, 150, 20, and 16. This neuron combination is the best in our fully connected layer combination test.

Second, we test various initialization methods. In our model, we find that Xavier is the best for the weight parameters where the initial value of the bias is set to 0. Third, we use LeakyReLU for activation function in fully connected layers rather than ReLU. In ReLU, if a negative number enters the input, then the output becomes 0. Thus, the neurons become inactive. LeakyReLU sets the hyperparameter α and multiplies the input by α to solve this problem. LeakyReLU can be calculated by Expression (8), and Fig. 7 shows the function. We set $\alpha$ to 0.1.

$$f_\alpha = \max(\alpha x, x) \tag{8}$$

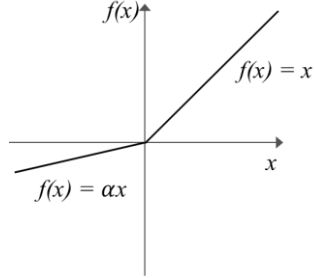

Fig. 7. Leaky ReLU

## 3.3 3D Grad-CAM

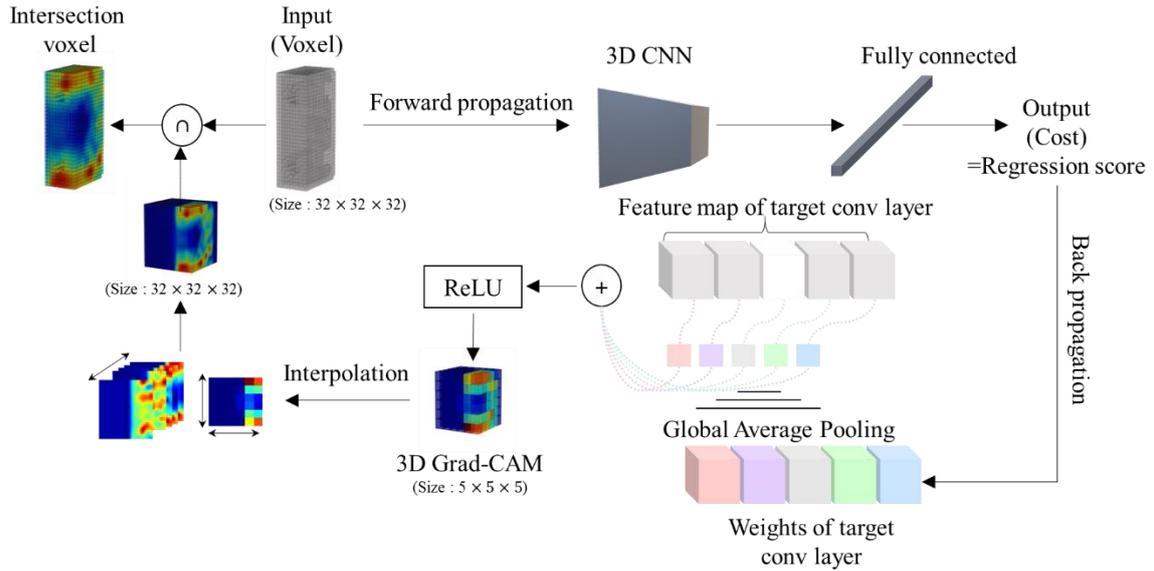

Fig. 8. 3D Grad-CAM process

The trained 3D CNN can be used to predict the cost of new parts, but the model is treated as a black box. The interpretability of the prediction result is essential to ensure reliability in the industry. This sub-section describes the implementation process of 3D Grad-CAM, which is the extended method of 2D Grad-CAM in Selvaraju et al. (2016), to visualize the 3D features. The process is organized in Fig. 8, and the equation is given as follows.

$$\alpha_l = \frac{1}{Z} \times \sum_i \sum_j \sum_k \frac{\partial y}{\partial A^l_{i,j,k}} \tag{9}$$

$$L_{\text{3D Grad CAM}} = ReLU\left(\sum_l \alpha_l \times A^l\right) \tag{10}$$

The 3D gradient of the predicted result can be obtained by backpropagation from the regression score $y$ to the output $A^l_{i,j,k}$ of target convolutional layer, where $A^l_{i,j,k}$ is feature map corresponding to the $i$-row j-column on the $k$-th channel of the $l$-th feature map. To multiply the influence of the $l$-th feature map, we perform the global average pooling process for the 3D gradient. This part is expressed as $\alpha_l$ in Equation (9). After the obtained $\alpha_l$ is linearly combined with the feature map $A^l_{i,j,k}$, it passes through the activation function ReLU to create a 3D Grad-CAM. This part is expressed as $L_{\text{3D Grad CAM}}$ in Equation (10).

The 3D Grad-CAM requires post-processing for visualization. The 3D Grad-CAM that came through the ReLU is smaller than the input size. If the size of the feature map is 5, then the size of the 3D Grad-CAM is also 5. The 3D Grad-CAM should be as large as the input resolution to ensure comparison with the input. Image interpolation is usually used to solve this problem. For the voxel, it is interpolated about row $i$ and column $j$ of each channel $k$ and then interpolated for channel $k$. The intersection of voxel space between the processed 3D Grad-CAM and input can represent the area with important features that the model has learned based on data.

## 4. Results and Discussion

This section discusses manufacturing cost estimation results and performs 3D Grad-CAM to interpret and describe the predicted results. Section 4.1 tests 90 baseline models described in Section 3.2.1 and our proposed model described in Section 3.3.2. Section 4.2 describes the machining features and explains these features through the proposed model. Section 4.3 describes the difficulty levels in CNC machining processing and interprets them through the proposed model.

### 4.1 Prediction of Manufacturing Costs

A total of 1,006 data are used, where 80% are used for training, and 20% are used for testing. For performance evaluation of the model, the root MSE (RMSE) and the mean absolute percentage error (MAPE) are used. The results of best models for each architecture are shown in Table 3, while all results of 108 models are presented in Appendix A. Each model has different architecture, input data type, normalization type, and loss function.

Table 3. Comparison of best models of each architecture

| Model characteristics | | | | Performance | |
|---|---|---|---|---|---|
| Architecture | Input | Normalization | Loss function | RMSE | MAPE |
| VoxNet | Voxel(32)[1], Mat, Vol | Min–max | MAE | **2716.24*** | 29.76 |
| | | Log | MSLE | 3109.07 | **18.58** |
| PointNet | Point(2048)[3], Mat, Vol | Log | MSE | **3503.78** | **21.17** |
| Dering and Tucker. (2017) | Voxel(32), Mat, Vol | Min–max | MAE | **1615.86** | 21.09 |
| | | Log | MSE | 2014.87 | **12.93** |
| Williams et al. (2019) | Voxel(64)[2], Mat, Vol | Log | MAE | 2165.35 | 14.12 |
| Ning et al. (2020) | Voxel(64), Mat, Vol | Log | MSE | 1047.47 | **10.63** |
| | | Log | MSLE | **1002.55** | 16.30 |
| Proposed | Voxel(32), Mat, Vol | Min–max | MAE | **1233.06** | 20.58 |
| | | Log | MAE | 1290.41 | **8.76** |

[1]Voxel(32) represents the size of the voxel 32×32×32.
[2]Voxel(64) represents the size of the voxel 64×64×64.
[3]Point(2048) represents the 1×2048 size of the point cloud.

*Bold font represents the best results for each architecture.

The key observations obtained through the experimental results are summarized as follows.

*First, the proposed model has the best performance in MAPE while using a small number of parameters.*
Fig. 9 is a comparison of the best models. The triangles in the figure represent models with the lowest MAPE, and the circles representing models with the lowest RMSE for each architecture. The squares represent the lowest model for MAPE and RMSE. For MAPE, the proposed model with log normalization and MAE loss function shows the best performance with a value of 8.76%. For RMSE, Ning et al. (2020) with log normalization and MSLE loss function shows the best performance with a value of 1002.55. However, in terms of number of parameters, the proposed model is approximately 20 times smaller than Ning et al. (2020). The proposed model has 4,245,369 parameters, while Ning et al. (2020) has 81,862,691 parameters. For training the model, the proposed model takes about 20.5 min using the GPU TiTAN Xp, while Ning's model takes about 6.7 h, which is nearly 20 times more than the proposed model. This finding shows that our model is more efficient while keeping good prediction performance.

Comparison of our model with Dering and Tucker. (2017)'s model as our benchmarking architecture shows that our model's RMSE and MAPE are 36% and 32% lower than those of Dering and Tucker. (2017)'s model. The optimal hyperparameter setting, the deeper fully connected layer design, and the extraction of more nonlinearity contribute to improving the performance of our model.

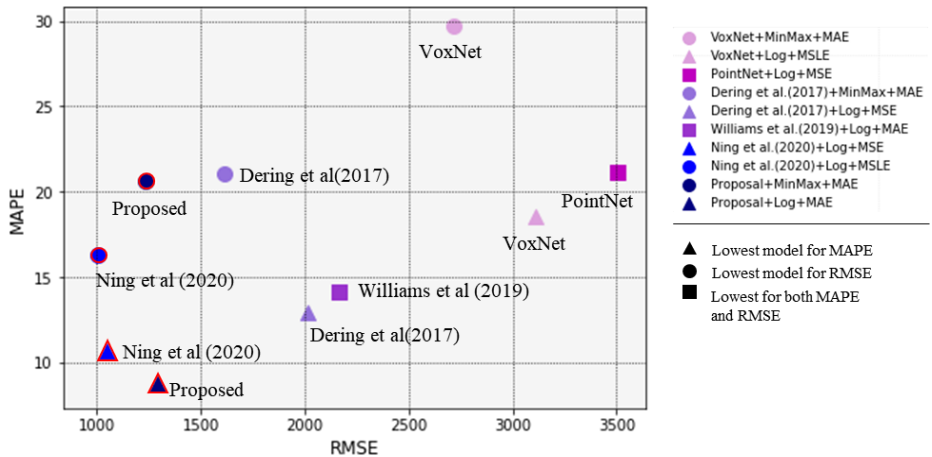

Fig. 9. Comparison of the best model for each architecture

Fig. 10 shows correlation between ground truths and predicted values for the best model of each architecture with lowest MAPE. As observed, the proposed model has the highest R value.

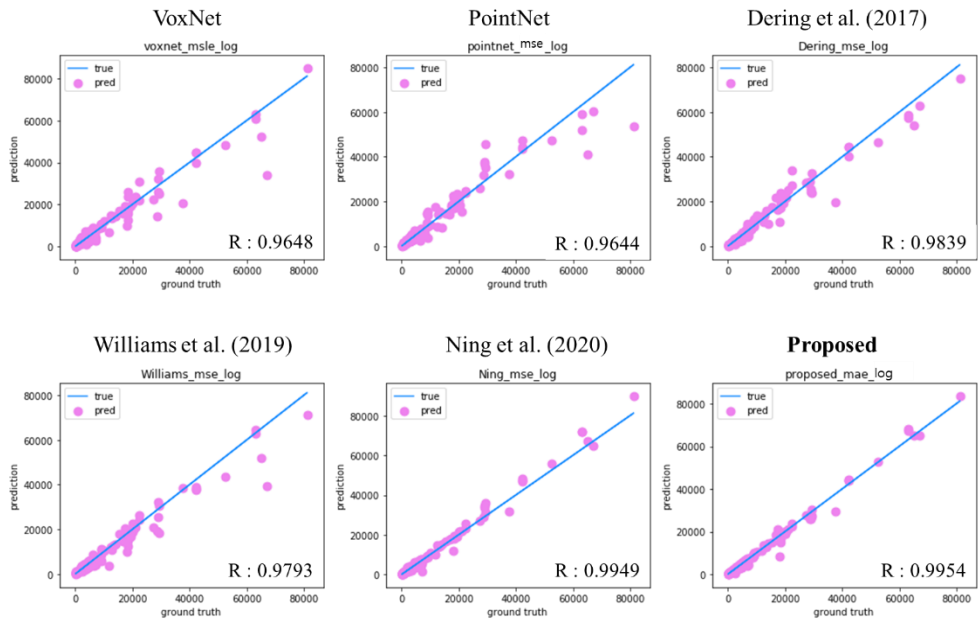

Fig. 10. Correlation between ground truths and predicted values

*Second, material data are essential for predicting manufacturing costs.*

Obviously, when input data include all CAD, material, and volume data, estimation performance increases in all models. In our case, material data considerably affect the estimation performance compared with volume data. Fig. 11 shows the evaluation result of the proposed model according to the input data type, where normalization and loss function types are log and MAE, respectively. Compared with voxel data, the accuracy improvement is greatest when material information is combined with voxel data. Specifically, RMSE decreases by 81.9% (10,644.07 → 1,925.68) and MAPE decreases by 50.1% (42.76 → 21.33). Adding volume information to voxel data has relatively little improvement in accuracy compared with adding material information. In particular, RMSE decreases by 13.1% (10,644.07 → 9,249.08) and MAPE decreases by 3.8% (42.76 → 41.12). Moreover, adding volume to voxel and material information considerably improves accuracy. Particularly, RMSE decreases by 33.0% (1,925.68 → 1,290.41) and MAPE decreases by 58.9% (21.33 → 8.76).

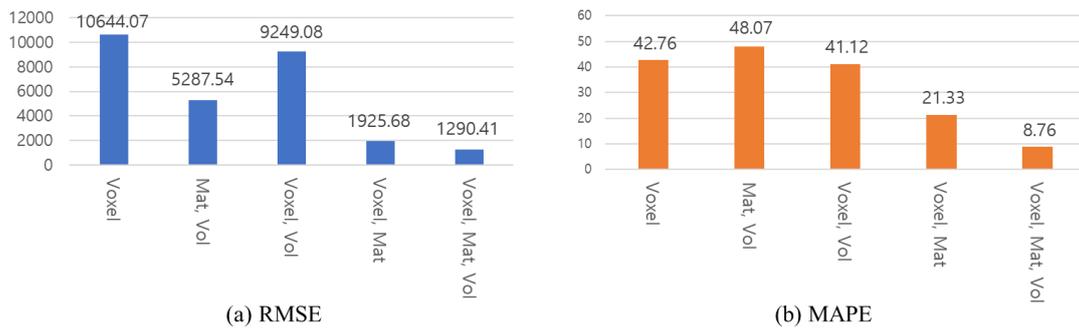

(a) RMSE  (b) MAPE

Fig. 11. Proposed model performance according to input data types

Ning et al. (2020) used only 3D CAD inputs without considering materials and volume. In our experiment, cost cannot be accurately predicted using only 3D CAD data. By adding material and volume information to Ning's model, predictive performance becomes much better.

The model consisting only of material and volume data performs considerably better than the model consisting of only voxel data in RMSE but slightly poorer in MAPE. The reason is that RMSE is sensitive to the size of the target value. Thus, it is sensitive to cost estimation of 3D data with large volumes, and the volume information becomes more important than the voxel shape. Meanwhile, MAPE focuses on the 3D shape (voxel) as much as the volume given that it is insensitive to the size of the target value.

*Third, normalization and loss function types depend on the architecture.*

Fig. 12 compares all models using 3D data, material, and volume. In the case of RMSE, the superiority of the normalization method cannot be distinguished. However, for MAPE, most models show that log normalization is considerably better than min–max normalization. The CAD data used for training consist of a large number of inexpensive parts and a small number of expensive parts. Therefore, we can train detailed cost differences between CAD data well after scaling the input data to be normally distributed through log normalization. As a result, the MAPE value is greatly improved when using log normalization.

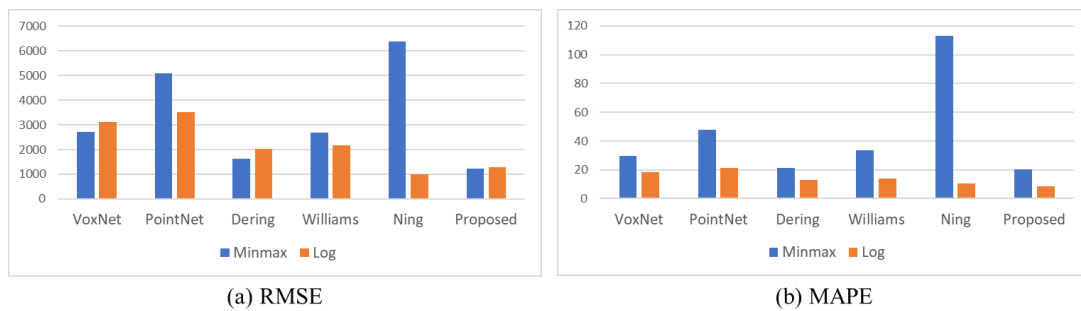

(a) RMSE  (b) MAPE

Fig. 12. Performance comparison according to normalization type

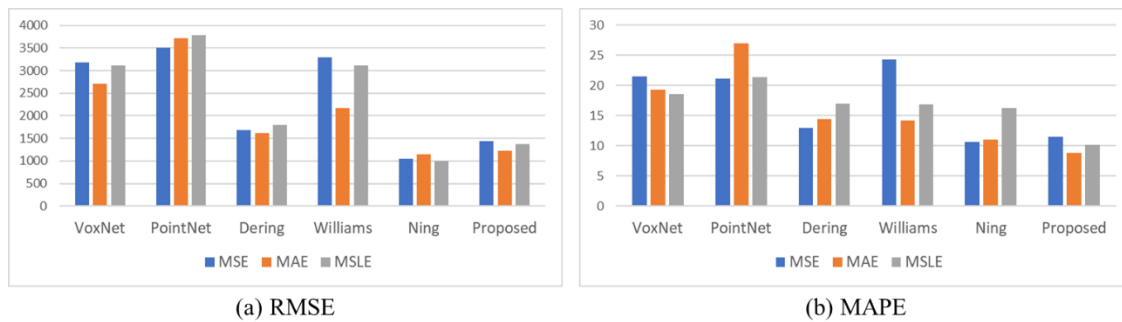

(a) RMSE  (b) MAPE

Fig. 13. Performance comparison according to loss function types

Fig. 13 compares all models according to loss function types. In the case of the loss function, the performance varies depending on the model. The loss function of the proposed model should also be used as MAE to minimize MAPE.

## 4.2. Explanation of Machining features

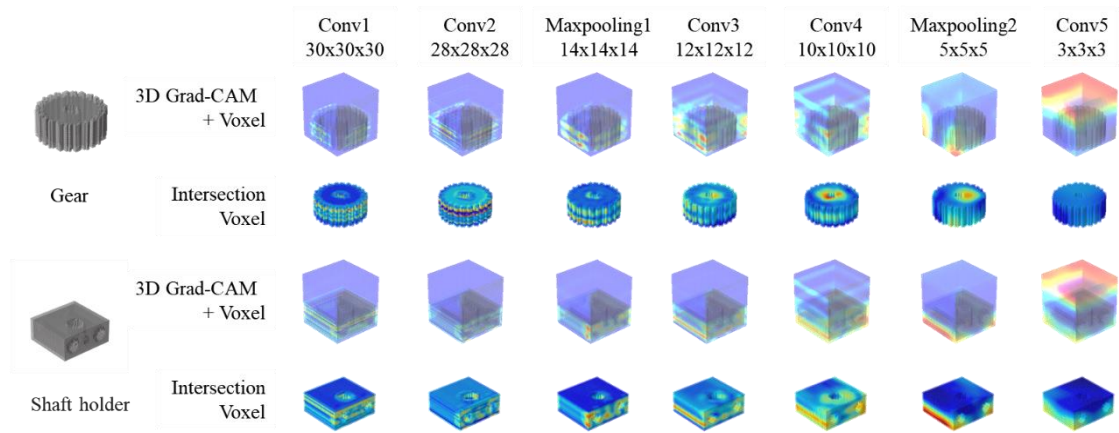

Fig. 14. 3D Grad-CAM visualization examples

Fig. 14 shows an example of Grad-CAM visualization of each feature map in the proposed model. The first row shows 3D Grad-CAM and the input voxel data, while the second row shows the intersection between 3D Grad-CAM and input voxel data. In the gear, the teeth and holes are mainly red highlighted. In the shaft holder, the area around the hole is highlighted in red. This condition means that these features considerably affect manufacturing costs. Fig. 14 shows differences in the degree to which features are clearly visible depending on the layer. For example, in Maxpooling1, machining features, such as holes, edges, and teeth, are well highlighted. In deeper layers, such as Conv5, the resolution is lowered, and the highlighted areas do not properly represent the important features.

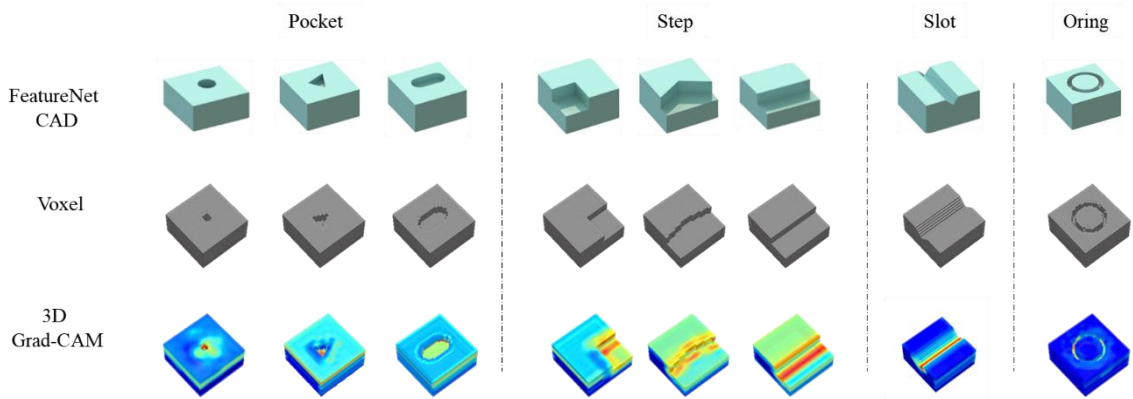

Fig. 15. CNC machining feature detection test

Zhang et al. (2018) defined CNC machining features as shown in the first row of Fig. 15. We develop CAD models based on each CNC machining feature and use them as test data to determine the capability of our proposed model to detect these machining features. As shown in the third row of Fig. 15, the 3D Grad-CAM with our proposed model highlights the machining features accurately. Fig. 16 shows some examples of test data with 3D Grad-CAM, which have defined CNC machining features. Notably, our proposed model generally detects these features well, although we do not give any information about CNC machining features to the model directly.

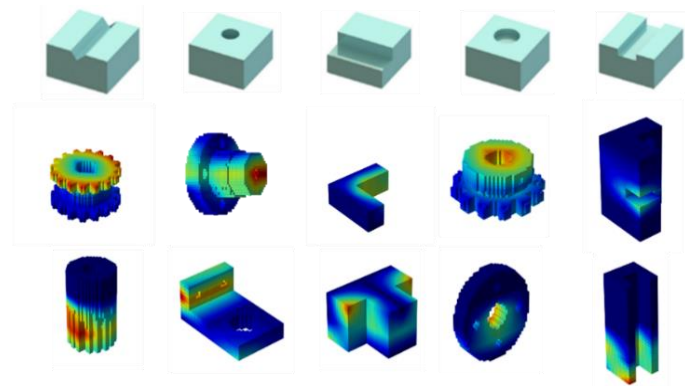

Fig. 16. CNC machining features in dataset

3D Grad-CAM allows us to visualize cost-critical areas in CAD model. Therefore, engineering designers in the early stages of product development can understand which parts of conceptual designs should be refined to reduce manufacturing costs.

## 4.3. Explanation of the Difficulty Level of Machining Processing

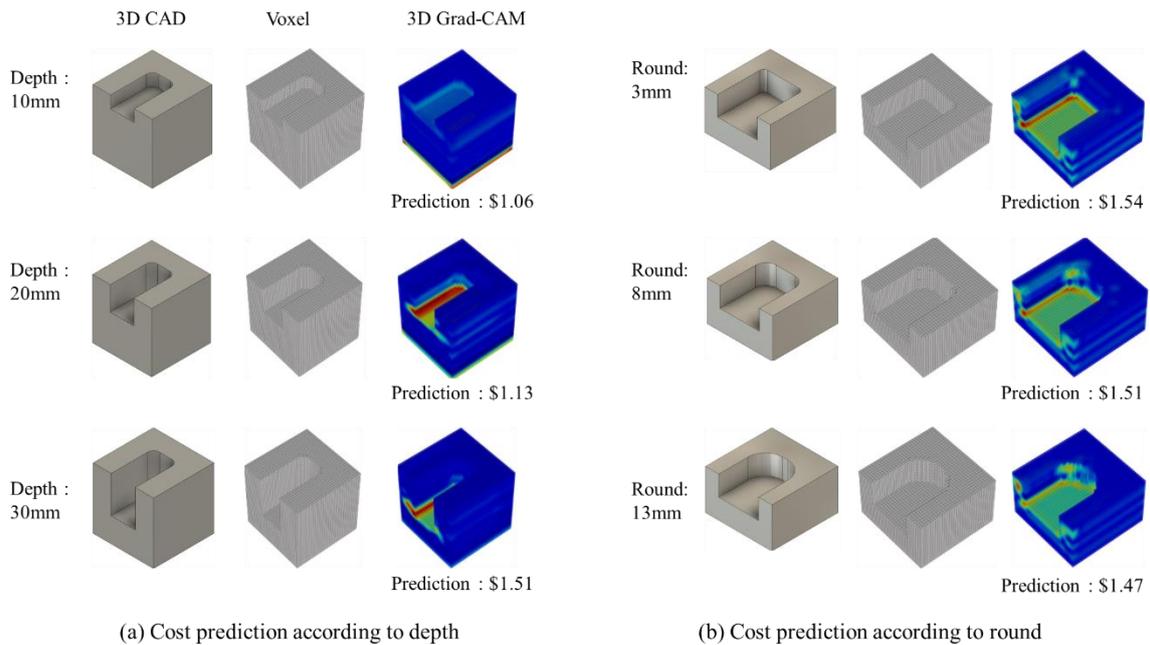

(a) Cost prediction according to depth        (b) Cost prediction according to round

Fig. 17. Recognition of difficulty of machining processing

CNC manufacturing costs are mainly related to machining difficulty, where machining is a process using a cutting tool. Cutting tools are divided into two main types: drills used for drilling holes and end mill used for cutting sidewalls. This subsection shows the results of an experiment to determine if the difference in the machining difficulty can be distinguished even in similar CAD shapes. The example in Fig. 17 shows the two cases of machining with end mill cutting.

The dataset for the experiment is generated by changing only certain parameters. The features used in Fig. 17(a) are designed with only a different degree of depth, each with a depth of 10, 20, and 30 mm. In reality, the cost is higher when the the processing depth is deeper. The reason is related to the operation of the end mill, which is well known as a tool used to cut axially for cutting the side of a workpiece. The vibration increases and the stability decreases when the depth of the process to be cut is deep. This condition is the cause of increased difficulty in processing. The prediction results of our model also show that the cost is higher when the processing depth is deeper. The 3D Grad-CAM also identifies areas that affect cost forecasting, and the floor surface is more concentratedly marked as an activation area when the workpiece depth is deeper.

The dataset used in Fig. 17(b) is generated differently for the round size of the corner. Round values for each corner are designed as 3, 8, and 13 mm. The large round means that it can be processed with such a large diameter end mill, which reduces processing time, and the processing stability is greater and the difficulty is lower when the diameter is larger. Prediction results of our model also show that that the cost is higher when the round size is smaller. The 3D Grad-CAM confirms that the round area of the corner is highlighted.

Experimental results have proven that our model can distinguish machining difficulty between similarly shaped CAD models. Our model can learn the effect of feature types and detail differences in the same feature on manufacturing costs based solely on CAD data without domain knowledge of processing difficulty.

## 5. Conclusion

This study proposes an explainable manufacturing cost estimation process for 3D CAD models using 3D-Grad CAM. Fig. 18 shows the usage scenario of the proposed process. Designers of customers create 3D CAD and input 3D models into the proposed model. After the CAD is automatically converted to voxels, the model visually identifies areas that considerably affect manufacturing costs. Then, designers or customers modify CAD to reduce manufacturing costs while satisfying engineering performance or design criteria.

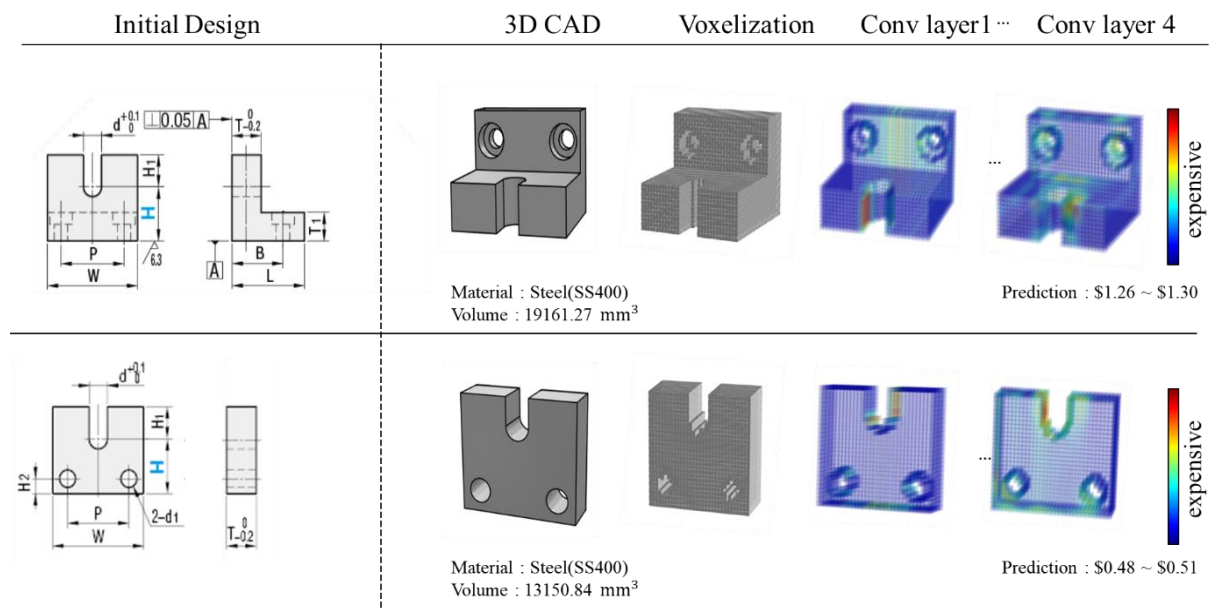

Fig. 18. Service scenario of the proposed process

The contributions of this study are as follows. First, our model tells designers which parts in 3D CAD need to be modified to reduce manufacturing costs. This feature increases the efficiency of redesign work for designers who lack knowledge regarding manufacturing cost. Previous studies have predicted the cost of manufacturing for designers, but designers had to decide for themselves where to modify.

Second, this study shows that CNC machining features in 3D CAD models can be detected only through manufacturing cost learning without direct learning of the machining features. If the machining features were trained directly through object detection or classification, as in previous studies, it would have required a substantial amount of labeling work (Zhang et al., 2018).

Third, our research has demonstrated that our data-driven model can learn the domain knowledge of CNC machining, such as the difficulty level of the machining process. Previous research has focused only on the accuracy of the predictions and has not tested whether domain knowledge is reflected in the deep learning model (Ning et al., 2020).

Lastly, this study provides some guidelines for building cost estimation models. For instance, previous work only uses 3D CAD data (Ning et al., 2020), but material and volume were found to be essential input data. In addition, converting 3D CAD into voxel is more suitable in terms of model accuracy and convergence than converting to point clouds. Furthermore, the choice of preprocessing method and optimizer depends on the deep learning architecture.

Future research will use CNC machining simulations and interviews with CNC experts to determine whether the features extracted through deep learning considerably affect the actual manufacturing cost and build a deep learning model that reflects them. Visualization and comparison will also be performed by introducing various XAI technologies. We also plan to add features that determine manufacturability.

**Declaration of Competing Interests**
The authors declare that they have no known competing financial interests or personal relationships that could have influenced the work reported in this study.

**Acknowledgments**
This study was supported by National Research Foundation of Korea (NRF) grants funded by the Korean government (MSIT) (No. 2017R1C1B2005266 and 2018R1A5A7025409) and HPC Support Project of the Ministry of Science and ICT and NIPA. The authors would like to thank San Ko of Ateam Ventures for his advice and ideas.

# Appendix A

**Table A1. Evaluation of overall model performance**

| Architecture | Input | Normalization type | Performance | | | | | |
|---|---|---|---|---|---|---|---|---|
| | | | RMSE | | | MAPE | | |
| | | | Loss function MSE | Loss function MAE | Loss function MSLE | Loss function MSE | Loss function MAE | Loss function MSLE |
| VoxNet | Voxel(32)[1] | Minmax | 9562.98 | 9643.55 | 9577.47 | 73.87 | 62.87 | 101.64 |
| | | Log | 9603.53 | 9792.50 | 10121.68 | 54.05 | 56.53 | 59.73 |
| | Voxel(32), Mat | Minmax | 2914.82 | 2725.73 | 2947.66 | 33.31 | 33.08 | 67.21 |
| | | Log | 3759.66 | 3384.20 | 3575.19 | 21.67 | 20.65 | 22.79 |
| | Voxel(32), Mat, Vol | Minmax | 3175.89 | **2716.24** | 3141.18 | 46.63 | 29.76 | 118.64 |
| | | Log | 4279.03 | 3582.26 | 3109.07 | 21.54 | 19.34 | **18.58** |
| PointNet | Point(2048) | Minmax | 8238.97 | 8301.45 | - | 73.85 | 47.8 | - |
| | | Log | 7848.32 | 8102.08 | - | 38.47 | 38.96 | - |
| | Point(2048), Mat | Minmax | - | 5165.62 | - | - | 40.05 | - |
| | | Log | 14360.73 | 8116.47 | - | 79.14 | 32.51 | - |
| | Point(2048), Mat, Vol | Minmax | 9430.96 | 5103.17 | - | 244.82 | 47.74 | - |
| | | Log | **3503.78** | 3724.84 | 3785.82 | **21.17** | 26.95 | 21.42 |
| Dering and Tucker. (2017) | Voxel(32) | Minmax | 9693.04 | 9493.96 | 9320.58 | 66.13 | 48.23 | 72.97 |
| | | Log | 9479.5 | 11310.82 | 10403.58 | 47.33 | 42.00 | 41.95 |
| | Voxel(32), Mat | Minmax | 2217.01 | 1813.02 | 2146.88 | 34.00 | 23.68 | 34.78 |
| | | Log | 2436.63 | 4176.82 | 4568.14 | 21.06 | 24.39 | 28.05 |
| | Voxel(32), Mat, Vol | Minmax | 1684.68 | **1615.86** | 1794.75 | 28.11 | 21.09 | 28.71 |
| | | Log | 2014.87 | 2574.63 | 3779.48 | **12.93** | 14.44 | 17.03 |
| Williams et al. (2019) | Voxel(64)[2] | Minmax | 10100.41 | 9599.6 | 9981.19 | 72.96 | 69.94 | 83.13 |
| | | Log | 9790.24 | 9842.57 | 9960.12 | 59.24 | 56.36 | 47.16 |
| | Voxel(64), Mat | Minmax | 3154.01 | 2932.19 | 3050.7 | 47.61 | 52.86 | 113.75 |
| | | Log | 2348.79 | 2771.57 | 2749.43 | 17.8 | 16.38 | 18.84 |
| | Voxel(64), Mat, Vol | Minmax | 3291 | 2695.29 | 3376.64 | 37.05 | 33.76 | 109.95 |
| | | Log | 3318.75 | **2165.35** | 3117.8 | 24.27 | **14.12** | 16.88 |
| Ning et al. (2020) | Voxel(64) | Minmax | 7102.95 | 7988.06 | 7788.32 | 78.11 | 59.81 | 68.10 |
| | | Log | 8626.35 | 8700.24 | 8789.09 | 44.88 | 42.5 | 40.75 |
| | Voxel(64), Mat | Minmax | 6915.51 | 7169.04 | 7097.48 | 134.8 | 115.98 | 123.73 |
| | | Log | 1565.27 | 1486.90 | 1299.94 | 12.86 | 11.43 | 10.93 |
| | Voxel(64), Mat, Vol | Minmax | 6904.21 | 7136.49 | 6374.03 | 139.1 | 113.18 | 115.15 |
| | | Log | 1047.47 | 1140.85 | **1002.55** | **10.63** | 10.96 | 16.30 |
| Proposed model | Voxel(32) | Log | 10290.33 | 10644.07 | 11151.45 | 45.29 | 42.76 | 41.18 |
| | Mat, Vol | Log | 5163.67 | 5287.54 | 5219.51 | 48.1 | 48.07 | 49.71 |
| | Voxel(32), Vol | Minmax | 7948.82 | 8271.09 | 7646.84 | 68.57 | 63.60 | 76.17 |
| | | Log | 8828.9 | 9249.08 | 8991.28 | 41.46 | 41.12 | 38.96 |
| | Voxel(32), Mat | Minmax | 1332.49 | 1377.27 | 1399.76 | 27.93 | 22.76 | 26.98 |
| | | Log | 2019.83 | 1925.68 | 1664.60 | 19.32 | 21.33 | 19.30 |
| | Voxel(32), Mat, Vol | Minmax | 1437.45 | **1233.06** | 1368.52 | 21.65 | 20.58 | 24.19 |
| | | Log | 1825.31 | 1290.41 | 1816.94 | 11.44 | **8.76** | 10.20 |

[1]Voxel (32) represents the size of the voxel 32x32x32.
[2]Voxel(64) represents the size of the voxel 64x64x64.
[3]Point(2048) represents the 1x2048 size of the point cloud.
[4]Bold font represents the best results for each architecture.
[5]All models using Voxel converged well, but some models using point cloud did not converge, so these models left the performance results blank.